\documentclass[12pt]{article}
\usepackage{amsmath, amsfonts, amssymb, amsthm, graphicx, geometry, arydshln, umoline, subfig, newtxtext, newtxmath, color, comment, soul, enumerate, bm}
\usepackage{appendix}
\usepackage{tikz}
\usetikzlibrary{positioning}
\usepackage{natbib}
\usepackage[colorlinks=true,citecolor=blue]{hyperref}
\usepackage{booktabs}
\usepackage{makecell}
\usepackage{bbm}

\theoremstyle{definition}
\newtheorem{theorem}{Theorem}

\newtheorem{corollary}{Corollary}

\newtheorem{definition}{Definition}
\newtheorem{example}{Example}

\newtheorem{proposition}{Proposition}

\begin{document}

\title{Safe manipulation of tops-only voting rules \thanks{The author is grateful to Takashi Kurihara and Satoshi Nakada for their helpful comments.}}
\author{
Toshiya Yoshimura\thanks{Undergraduate School of Management, Department of Business Economics, Tokyo University of Science, 1-11-2, Fujimi, Chiyoda-ku, Tokyo, 102-0071, Japan. Email: tossyyossy01@gmail.com}
}
\date{\today}
\maketitle
\begin{abstract}

We study tops-only voting rules on the domain of all preferences with a unique top, focusing on manipulation by risk-avoiding voters.
Following \citet{Hartman2025}, we measure resistance to manipulation by the RAT-degree: the minimum number of other voters' reports that a voter needs to observe for a ``safe'' manipulation to become possible.
Given the observed reports, such a manipulation weakly improves on truthful reporting under every completion of the unobserved reports and strictly improves on it under at least one completion.
For any finite set of at least three alternatives, we show that anonymity, unanimity, and winner monotonicity imply an upper bound of $\lceil n/2\rceil$ on the RAT-degree, where $n$ is the number of voters.
We provide exact characterizations of the RAT-degree for anonymous generalized median voter schemes and voting by quotas, without imposing single-peakedness or separability on preferences.
For odd $n$, the median voter rule and objectwise majority rule uniquely maximize the RAT-degree within their respective classes and attain the general upper bound.

\end{abstract}

\section{Introduction}
Strategy-proofness requires that truthful reporting be a weakly dominant strategy for every voter.
This property may be a demanding requirement: a single profitable deviation at a single preference profile is enough to violate it.
The impossibility results of \citet{Gibbard1973,Satterthwaite1975} show that, on an unrestricted preference domain with at least three alternatives, every onto and strategy-proof social choice function is dictatorial.
However, the existence of a profitable deviation does not by itself imply that a voter can safely exploit it.
When other voters' preferences are not fully known, a deviation that improves the outcome at one profile may yield a worse outcome than truthful reporting at another.
As \citet{Klamler2014} emphasizes, even small errors in a voter's information about others' preferences can turn an intended gain from manipulation into a loss.
These considerations motivate evaluating voting rules in terms of the information required to manipulate them without risking a loss.

\citet{Bu2023} propose a relaxation of truthfulness in cake cutting that excludes deviations that are sometimes beneficial and never harmful.
Following \citet{Hartman2025}, we refer to this property as risk-avoiding truthfulness (RAT).
To account for partial information, suppose that an agent knows the reports of a subset of other voters.
A safe manipulation makes her weakly better off than truthful reporting for every profile of the unknown reports and strictly better off for at least one such profile, holding the known reports fixed.
\citet{Hartman2025} introduce the RAT-degree as the smallest number of other agents whose reports, if known, can permit such a manipulation, assigning the total number of agents to strategy-proof mechanisms.
Therefore, a higher RAT-degree indicates that more information is needed before a profitable deviation can be undertaken without risking a loss.
This approach compares voting rules in terms of the amount of information required for safe manipulation, without specifying beliefs or a probability distribution over the unknown reports.

Tops-only voting provides a particularly natural setting for the known-agents framework of \citet{Hartman2025}, which assumes exact knowledge of some other agents' reports.
Knowing another agent's complete preference ranking or valuation function may require detailed information about her preferences.
Under a tops-only rule, each report consists of a single alternative, so knowing another voter's report requires only knowing which alternative she submits.
Such information may be available through discussions or public declarations before voting, even when complete preference rankings remain unknown.
\citet{DryzekList2003} discuss preference revelation during deliberation, while \citet{Nurmi2023} points out that such revelation may create opportunities for strategic misrepresentation.
Moreover, \citet{Klamler2014} suggests that voters may be better informed about others' first choices than about their rankings of less preferred alternatives.
Motivated by these considerations, we study tops-only rules for $n\geq2$ voters choosing from a finite set of at least three alternatives, allowing all preferences with a unique top, including indifference below it.
A potential manipulator knows the reports of a subset of other voters, while every completion of the remaining reports is regarded as possible.

On the domain of strict preferences, \citet{MausPetersStorcken2007} show that, when $n>|X|\geq3$, the unanimity rule with a status quo minimizes the number of manipulable preference profiles among anonymous, surjective tops-only rules.
Nevertheless, its RAT-degree is one for every $n\geq2$: a single observed report can suffice for safe manipulation.
The rule's tendency to preserve the status quo limits the profiles at which deviations can be profitable and can also eliminate their downside risk, since an observed ballot can reveal that voting for a preferred compromise either secures it or leaves in place the status quo that truthful reporting would yield. 
Minimizing the number of manipulation opportunities therefore need not maximize the information required to exploit them safely.
This example illustrates that focusing on safe manipulation can yield conclusions about a voting rule's resistance to strategic manipulation that differ from those suggested by existing criteria.

Our first main result shows that every anonymous and unanimous tops-only voting rule satisfying winner monotonicity has RAT-degree at most $\lceil n/2\rceil$ (Theorem~\ref{thm:rat-upper-bound}).
Winner monotonicity requires the current winner to remain chosen when a voter switches her report to that alternative.
The bound means that every rule in this class admits a safe manipulation in some situation where at most $\lceil n/2\rceil$ other reports are known.
Anonymity and unanimity exclude dictatorships and constant rules, respectively, both of which are strategy-proof and have RAT-degree $n$.

We apply the RAT-degree to two canonical classes of tops-only voting rules: anonymous generalized median voter schemes and voting by quotas.
\citet{Moulin1980} establishes the strategy-proofness of generalized median voter schemes under single-peaked preferences, while \citet{Weymark2011} develops the associated characterizations for arbitrary subsets of the real line, including finite ordered sets.
\citet{Barbera1991} establish the strategy-proofness of voting by committees under separable preferences, with voting by quotas forming the anonymous subclass.
More generally, \citet{LeBretonSen1999} show that strategy-proofness requires decomposability across components on sufficiently rich domains of strict separable preferences.
These results provide natural benchmarks, but the rules remain well defined when voters' preferences fail to satisfy single-peakedness or separability.
Our setting is closely related to \citet{ArribillagaMasso2016,ArribillagaMasso2017}, who compare the manipulability of these families, and to \citet{ArribillagaBonifacio2024}, who analyze their obvious manipulations.
Since voters in actual elections need not have perfectly single-peaked or separable preferences, we ask which rules within each class require the most information before a safe manipulation becomes possible.

For an anonymous generalized median voter scheme $f^\alpha$ with $n-1$ fixed ballots, we establish $d_{\mathrm{RAT}}(f^\alpha)=1+\min\{\ell,r\}$, where $\ell$ and $r$ are the numbers of fixed ballots at the left and right endpoints, respectively.
For a voting by quotas rule $f^q$ on an object set $O$, we establish $d_{\mathrm{RAT}}(f^q)=\min_{a\in O}\min\{q_a,n-q_a+1\}$, where $q_a$ is the quota for including object $a$.
When $n$ is odd, these formulas identify the median voter rule and objectwise majority as the unique maximizers within their respective classes, each attaining RAT-degree $(n+1)/2=\lceil n/2\rceil$.
Since both classes satisfy unanimity, anonymity, and winner monotonicity, these two rules also maximize the RAT-degree among all tops-only rules satisfying these three axioms.
Both rules satisfy the majority criterion, so their attainment of the general upper bound also follows directly from the sufficient condition in Theorem~\ref{thm:optimal}.

Our results distinguish rules that are equally manipulable under the criterion of \citet{ArribillagaMasso2016,ArribillagaMasso2017}, which compares the sets of individual preferences permitting a profitable manipulation.
For odd $n$, the median voter rule and objectwise majority belong to the most manipulable equivalence classes within their respective families, yet uniquely maximize the RAT-degree.
This contrast separates the existence of profitable deviations from the information needed to exploit them safely.

We also relate our measure to non-obvious manipulability (NOM), introduced by \citet{TroyanMorrill2020} and studied for tops-only rules by \citet{ArribillagaBonifacio2024}.
NOM concerns improvements in the best or worst possible outcome, and neither NOM nor maximal RAT-degree ($\lceil n/2 \rceil)$ implies the other in general.
Within the anonymous generalized median class, a RAT-degree of at least two is sufficient for NOM; within voting by quotas, it is both necessary and sufficient.

Our analysis is related to the literature on strategic voting under incomplete information.
\citet{ConitzerWalshXia2011} represent a voter's information by a set of possible profiles and study dominating manipulations, which weakly improve upon truthful reporting at every profile in that set and strictly improve it at some profile.
\citet{ReijngoudEndriss2012,EndrissEtAl2016} examine how information about election winners, candidate scores, and majority comparisons affects manipulation and voting dynamics.
We adopt the known-agents framework of \citet{Hartman2025}, taking information to consist of the reports of a subset of voters.
Our contribution is to derive general bounds under voting axioms and exact information thresholds within canonical classes of tops-only rules.

A related approach is regret-free truth-telling, introduced by \citet{Fernandez2020} in matching and subsequently studied in school choice with consent by \citet{ChenMoller2024}.
This criterion evaluates deviations using the information inferred from an observed outcome.
Truthful reporting is regret-free if, after every possible observation, no alternative report is weakly better in all compatible scenarios and strictly better in at least one.
\citet{ArribillagaBonifacioFernandez2022} apply this approach to voting and show that regret-free truth-telling coincides with strategy-proofness for tops-only rules on the unrestricted domain of strict preferences.
Thus, their criterion does not distinguish among non-strategy-proof tops-only rules.
The RAT-degree provides a different comparison by measuring how many individual reports must be known before a safe manipulation becomes possible.

\section{Preliminaries}

\subsection{Model}

Let $N=\{1,\ldots,n\}$, $n\ge 2$, be a finite set of voters, and let $X,\ |X|\ge 3$, be a finite set of alternatives.
Each voter $i\in N$ has a complete and transitive preference relation $R_i$ over $X$.
Let $P_i$ denote the strict preference relation induced by $R_i$; that is, for all $x,y \in X$, $x P_i y$ if and only if $\lnot y R_i x$.
Similarly, let $I_i$ denote the indifference preference relation induced by $R_i$, namely, for all $x,y \in X$, $x I_i y$ if and only if $x R_i y  \land y R_i x$.
We also restrict our attention to preferences with a unique top.
Let $\mathcal R$ denote the set of all such preference relations, and let $R=(R_i)_{i\in N}\in\mathcal R^n$ denote a preference profile.
For each $R_i\in\mathcal R$, let $\tau(R_i)\in X$ denote its unique top. 
For a profile $R\in\mathcal R^n$, write $\tau(R)=(\tau(R_1),\ldots,\tau(R_n))$.
In this paper, we focus on tops-only rule.
Let $f:X^n \to X$ be a tops-only voting rule. 
Whenever the underlying preference profile $R$ is clear from the context, we simply write $t_i$ for $\tau(R_i)$ and $t$ for $\tau(R)$.

\subsection{RAT-degree}

We first recall the standard notion of strategy-proofness as a benchmark.

A tops-only voting rule $f$ is \emph{strategy-proof} if, for every $i\in N$, every $R=(R_i,R_{-i})\in\mathcal R^n$, and every $t_i'\in X$, $f(t_i,t_{-i})R_i f(t_i',t_{-i})$.
Strategy-proofness requires truthful reporting to be optimal for every realized profile of the other voters' preferences.
We now consider a voter who has only partial information about the reports of the other voters.

Fix a voter $i\in N$.
Let $K\subseteq N\setminus\{i\}$ denote the set of voters whose reports are known to voter $i$, and let $U=N\setminus(K\cup\{i\})$ denote the set of voters whose reports are unknown to voter $i$.
For a top profile $t\in X^n$, we write $t=(t_i,t_K,t_U)$, where $t_K=(t_j)_{j\in K}$ and $t_U=(t_j)_{j\in U}$.
Let $X^K$ and $X^U$ denote the set of all possible top profiles of the voters in $K$ and $U$.
Voter $i$ observes $t_K$ and considers every $t_U'\in X^U$ a possible realization of the unknown voters' reports.

Following \citet{Hartman2025}, we introduce the notions of safe manipulation and RAT-degree.

\begin{definition}
Fix a voter $i\in N$, a set $K\subseteq N\setminus\{i\}$, and let $U=N\setminus(K\cup\{i\})$.
For $R_i\in\mathcal R$, $t_i'\in X$, and $t_K\in X^K$,
the report $t_i'$ is \emph{profitable given $(R_i,t_K)$} if $f(t_i',t_K,t_U)P_i f(t_i,t_K,t_U)$ for some $t_U\in X^U$.
The report $t_i'$ is \emph{safe given $(R_i,t_K)$} if
$f(t_i',t_K,t_U)R_i f(t_i,t_K,t_U)$ for every $t_U\in X^U$.
The report $t_i'$ is a
\emph{safe manipulation given $(R_i,t_K)$}
if it is both profitable and safe
given $(R_i,t_K)$.
\end{definition}

A safe manipulation offers a possible improvement without any risk of making the voter worse off than under truthful reporting, given the observed reports $t_K$ and regardless of the unknown voters' reports.
Using this notion, we define when a rule is
$k$-known-voter safely manipulable.
\begin{definition}
For each $k\in\{0,\ldots,n-1\}$,
a tops-only voting rule $f$ is
\emph{$k$-known-voter safely manipulable}
if there exist $i\in N$,
$K\subseteq N\setminus\{i\}$ with $|K|=k$,
$R_i\in\mathcal R$, $t_i'\in X$,
and $t_K\in X^K$
such that $t_i'$ is a safe manipulation
given $(R_i,t_K)$.
\end{definition}

\citet[Proposition~3.3]{Hartman2025} show that, for every $k\in\{0,\ldots,n-2\}$, if $f$ is $k$-known-voter safely manipulable, then it is also $(k+1)$-known-voter safely manipulable.
This monotonicity motivates measuring resistance to safe manipulation by the minimum number of known reports that can permit it.
Accordingly, we define the RAT-degree as follows.

\begin{definition}
The \emph{RAT-degree} of a tops-only voting rule $f$, denoted by
$d_{\mathrm{RAT}}(f)$, is the minimum
$k\in\{0,\ldots,n-1\}$ for which
$f$ is $k$-known-voter safely manipulable,
or $n$ if no such $k$ exists.
\end{definition}

Thus, a larger RAT-degree means that more information about the other voters' preferences is required before a voter can manipulate the social choice function safely.
Thus, a voting rule with a lower RAT-degree is easier to manipulate safely.
Moreover, it is easy to verify that $d_{\mathrm{RAT}}(f)=n$ if and only if $f$ is strategy-proof.
Strategy-proofness is therefore a demanding requirement: it rules out safe manipulation even when a voter knows all other voters' reports.
A single profitable deviation at a single profile violates this property, even if the same deviation could make the voter worse off under other reports.

\begin{example}[Unanimity rule with status quo]
\label{ex:status-quo}
Fix $z\in X$.
The \emph{unanimity rule with status quo $z$}, denoted by $f^z:X^n\to X$, is defined by
\[
f^z(t)=
\begin{cases}
x, & \text{if } t_i=x \text{ for every } i\in N,\\
z, & \text{otherwise}.
\end{cases}
\]
On the domain of strict preferences, \citet{MausPetersStorcken2007} show that, when $n>|X|\geq3$, this rule minimizes the number of manipulable preference profiles among anonymous, surjective tops-only rules.
Nevertheless,
\[
d_{\mathrm{RAT}}(f^z)=1
\qquad\text{for every } n\geq2.
\]
Thus, a single observed report can suffice for safe manipulation, regardless of the size of the electorate.

To establish the lower bound, fix any voter $i\in N$ and any $R_i\in\mathcal R$, and let $p=\tau(R_i)$.
If $p=z$, truthful reporting always yields the voter's top, so no deviation is profitable.
If $p\neq z$, consider the completion in which every other voter reports $p$.
Truthful reporting yields $p$, whereas every misreport yields $z$, which is strictly worse.
Hence, no safe manipulation is possible without information, and $d_{\mathrm{RAT}}(f^z)\geq1$.

For the reverse inequality, choose distinct $p,q\in X\setminus\{z\}$ and a preference $R_i\in\mathcal R$ with $\tau(R_i)=p$ and $qP_i z$.
Choose $j\in N\setminus\{i\}$, let $K=\{j\}$, and fix the observed profile $t_K=(q)_{h\in K}$.
Let $U=N\setminus(K\cup\{i\})$.
For every $t_U\in X^U$, we have $f^z(p,t_K,t_U)=z$ and $f^z(q,t_K,t_U)\in\{q,z\}$.
Hence, $f^z(q,t_K,t_U)R_i f^z(p,t_K,t_U)$ for every $t_U\in X^U$.
At the completion $\bar t_U=(q)_{h\in U}$, we obtain $f^z(q,t_K,\bar t_U)=qP_i z=f^z(p,t_K,\bar t_U)$.
Therefore, reporting $q$ is a safe manipulation given $(R_i,t_K)$, establishing $d_{\mathrm{RAT}}(f^z)=1$.
The same argument applies when preferences are restricted to strict orders.

This example illustrates why few opportunities for manipulation need not imply strong resistance to safe manipulation.
The rule changes the status quo only under unanimity, which limits the circumstances in which a deviation can affect the outcome.
Yet, once a known report rules out unanimity at the voter's true top, matching that report can secure a preferred compromise while otherwise leaving the status quo unchanged.
The limited scope for changing the outcome can therefore also eliminate the downside risk of a deviation, making misreporting attractive even to a risk-avoiding voter.
By measuring the information needed to exploit manipulation opportunities safely, the RAT-degree reveals an aspect of strategic vulnerability that is not captured by counting manipulable profiles.
\end{example}

\section{A general upper bound}

In this section, we characterize an upper bound of RAT-degree of tops-only voting rules which satisfy simple axioms.

\subsection{Axioms}

First, we introduce a monotonicity property.
Positive responsiveness, as defined by \citet[Definition~2.8]{Muller2024}, requires that the chosen alternative remain chosen when voters raise it in their rankings without changing the relative rankings of the other alternatives.
For tops-only rules on our domain, this condition is equivalent to winner monotonicity: if a voter changes her reported top to the current winner, that alternative must remain the winner.
We require the voting rules considered here to satisfy this monotonicity property.
Formally, we define it as follows.

\paragraph{Winner monotonicity:}
A tops-only voting rule $f:X^n\to X$
satisfies \emph{winner monotonicity} (WM) if,
for every $t\in X^n$, every $i\in N$, and every $x\in X$,
\[
f(t)=x \ \Longrightarrow \ f(x,t_{-i})=x.
\]

Next, we introduce the two axioms, anonymity and unanimity.
Both axioms are standard requirements in social choice theory.
Anonymity requires that all voters be treated equally.
Unanimity requires that an alternative be selected whenever all voters report that alternative.
\paragraph{Anonymity:}
A tops-only voting rule $f:X^n\to X$
satisfies \emph{anonymity} (A) if, for every $t\in X^n$
and every permutation $\pi:N\to N$, $f(t_1,\ldots,t_n) = f(t_{\pi(1)},\ldots,t_{\pi(n)}).$

\paragraph{Unanimity:}
A tops-only voting rule $f:X^n\to X$ satisfies \emph{unanimity} (U) if, for every $x\in X$, $f(x,\ldots,x)=x.$

\subsection{The upper bound}
The following theorem establishes an upper bound on the RAT-degree of tops-only voting rules satisfying winner monotonicity, anonymity, and unanimity. 
Under these three axioms, the RAT-degree cannot exceed about half the number of voters.

\begin{theorem}
\label{thm:rat-upper-bound}
If a tops-only voting rule $f:X^n\to X$
satisfies anonymity, unanimity and winner monotonicity, then
\[
d_{\mathrm{RAT}}(f)
\le
\left\lceil \frac{n}{2}\right\rceil.
\]

\end{theorem}

Theorem~\ref{thm:rat-upper-bound} has two implications.
First, it gives a negative result: every tops-only rule satisfying these mild axioms admits a safe manipulation in some situation where at most $\lceil n/2\rceil$ other reports are known.
Thus, no rule satisfying these requirements can entirely avoid situations in which knowledge of only about half of the electorate's reports permits a profitable deviation without any risk of a loss.

Second, the theorem provides a benchmark for optimality.
Any rule satisfying the same axioms and attaining
$d_{\mathrm{RAT}}(f)=\lceil n/2\rceil$
maximizes the minimum amount of information needed for safe manipulation within this class.
Such a rule rules out safe manipulation whenever fewer than $\lceil n/2\rceil$ other reports are known, regardless of the voter's preference or the observed reports.
Since no rule satisfying the three axioms can raise this threshold further, attaining the upper bound represents the strongest resistance to safe manipulation compatible with these requirements.
As we show below, this bound is attainable.
The theorem therefore both establishes a limit on resistance to safe manipulation and supplies a criterion for identifying rules that are optimal in this respect.

If any one of the three axioms is dropped, there exist tops-only voting rules with RAT-degrees exceeding the upper bound established in Theorem \ref{thm:rat-upper-bound}. 
We illustrate this observation with the following examples.

\paragraph{(A):}
There exists $i^D \in N$ such that for any $t \in X^n$,
\[
f^D(t) = t_{i^D}.
\]
This rule satisfies all axioms in Theorem \ref{thm:rat-upper-bound} except for anonymity.
This rule is known as a dictatorship and is clearly strategy-proof.
Therefore, $d_{\mathrm{RAT}}(f^D)=n.$

\paragraph{(U):}  There exists $c \in X$ such that for any $t \in X^n$,
\[
f^C(t) = c.
\]
This rule satisfies all axioms in Theorem \ref{thm:rat-upper-bound} except for unanimity.
This rule is known as a constant rule and is clearly strategy-proof.
Therefore, $d_{\mathrm{RAT}}(f^C)=n.$

\paragraph{(WM):} 
Let $X=\{0,1,2\}$.
For each top profile $t=(t_1,\ldots,t_n)\in X^n$, define the tops-only rule $f^{\mathrm{mod}}$ as follows.
If $t_1=\cdots=t_n=x$ for some $x\in X$, let $f^{\mathrm{mod}}(t)=x$.
Otherwise, let $f^{\mathrm{mod}}(t)$ be the unique element of $X$ satisfying
$f^{\mathrm{mod}}(t)\equiv\sum_{i\in N}t_i\pmod 3$.

By construction, $f^{\mathrm{mod}}$ is anonymous and unanimous.
For every $n\ge 4$, $d_{\mathrm{RAT}}(f^{\mathrm{mod}})=n-1$.

The intuition is simple. As long as at least one other voter's report remains unknown, that voter can report an alternative such that the sum of all other voters' reports is congruent to zero modulo $3$.
In this case, the outcome coincides with voter $i$'s own report. Since truthful reporting yields their most preferred alternative, any misreport results in a strictly worse outcome.
Thus, no manipulation can be safe unless all other voters' reports are known.

Theorem~\ref{thm:rat-upper-bound} establishes an upper bound on the RAT-degree of rules satisfying anonymity, unanimity, and winner monotonicity.
A natural next question is how to characterize the rules that attain this bound.
The following example shows that this class includes rules that select an alternative even when it loses a pairwise majority comparison.

\begin{example}
\label{counter_ex_thm2}
Let $N=\{1,2,3\}$ and $X=\{a,b,c\}$.
For each $t\in X^3$ and $x\in X$, let
$n_x(t)=|\{i\in N:t_i=x\}|$.
Define a tops-only voting rule $f:X^3\to X$ by
\[
f(t)=
\begin{cases}
b, & \text{if } n_b(t)\geq1,\\
a, & \text{if } n_b(t)=0 \text{ and } n_a(t)\geq2,\\
c, & \text{otherwise}.
\end{cases}
\]
This rule satisfies anonymity, unanimity, and winner monotonicity.
It selects $b$ whenever at least one voter reports $b$.
In particular, $f(a,a,b)=b$, even though two of the three voters prefer $a$ to $b$, so that $a$ defeats $b$ in a pairwise majority comparison..

By Theorem~\ref{thm:rat-upper-bound}, $d_{\mathrm{RAT}}(f)\leq2$, so it suffices to establish the reverse inequality.
Fix any voter $i\in N$, any preference $R_i\in\mathcal R$, and any observed profile $t_K$ with $|K|\leq1$.
Let $p=\tau(R_i)$ and $U=N\setminus(K\cup\{i\})$.
If a known report is $b$, the outcome is fixed at $b$, so no profitable deviation is possible.
Otherwise, choose a completion $\bar t_U$ in which the two voters other than $i$ report $a$ and $c$, one each.
Such a completion is consistent with the observed reports because at most one report is known and any known report belongs to $\{a,c\}$.
Under this completion,
\[
f(s_i,t_K,\bar t_U)=s_i
\qquad\text{for every }s_i\in X.
\]
Hence, every misreport $s_i\neq p$ satisfies
\[
f(p,t_K,\bar t_U)=p
\;P_i\;
s_i=f(s_i,t_K,\bar t_U).
\]
Thus, no safe manipulation is possible when at most one other report is known.
Consequently,
\[
d_{\mathrm{RAT}}(f)=2=\left\lceil\frac n2\right\rceil.
\]
Therefore, the rule therefore attains the upper bound.
\end{example}

Example~\ref{counter_ex_thm2} shows that maximal resistance to safe manipulation can coexist with selecting $b$ even when a majority of voters prefer $a$ to $b$.
A complete characterization of the rules attaining the upper bound must accommodate such choices.
This suggests that finding necessary and sufficient conditions with a simple and compelling normative interpretation may be difficult.
In Section~5.1, we discuss a sufficient condition for attaining the upper bound.

\section{Application}
Classical strategy-proofness results characterize broad classes of tops-only voting rules on restricted preference domains, including generalized median voter schemes under single-peakedness and voting by quotas under separability \citep{Moulin1980,Barbera1991}.
These characterizations rely on domain restrictions that voters' preferences need not satisfy in practice.
When the domain is enlarged to all preferences with a unique top, as in our setting, these rules remain well defined, but they are no longer strategy-proof.
Comparing their RAT-degrees nevertheless allows us to distinguish them according to the information required for safe manipulation.
This comparison provides a criterion for selecting among rules that are all strategy-proof on the relevant restricted domain but differ in how much information is needed for safe manipulation when the domain restrictions are removed.

\subsection{Generalized median voter schemes}
\subsubsection{Generalized median voter schemes}

We now restrict attention to anonymous generalized median voter schemes.
Let $X=\{x_1,\ldots,x_m\}$, where $m\ge 3$, be equipped with a fixed ordering $<$ such that $x_1<\cdots<x_m$.

Let $\alpha=(\alpha_1,\ldots,\alpha_{n-1})\in X^{n-1}$ be a vector of \emph{fixed ballots} satisfying $\alpha_1\le\cdots\le\alpha_{n-1}$.
The \emph{anonymous generalized median voter scheme} associated with $\alpha$ is the tops-only voting rule $f^\alpha: X^n\to X$ defined by
\[
f^\alpha(t)
=
\operatorname{med}(t_1,\ldots, t_n,\alpha_1,\ldots,\alpha_{n-1})
\]
Thus, the outcome is the median of the voters' reported top alternatives and the fixed ballots.
Since the outcome depends only on the multiset of reported top alternatives, $f^\alpha$ is anonymous.

For each vector of fixed ballots $\alpha$, let
\[
\ell(\alpha)
=
\bigl|\{j\in\{1,\ldots,n-1\}:\alpha_j=x_1\}\bigr|,
\qquad
r(\alpha)
=
\bigl|\{j\in\{1,\ldots,n-1\}:\alpha_j=x_m\}\bigr|.
\]
These quantities count the fixed ballots located at the left and right endpoints of $X$, respectively.

We next recall the classical connection between anonymous generalized median voter schemes and strategy-proofness on the single-peaked domain.
A preference relation $R_i\in\mathcal R$ is \emph{single-peaked} with respect to $<$ if, for every $x,y\in X$,
\[
x<y\le\tau(R_i)\ \Longrightarrow\ yP_i x,
\qquad
\tau(R_i)\le x<y\ \Longrightarrow\ xP_i y.
\]
Thus, as one moves away from voter $i$'s top alternative in either direction, alternatives become strictly less preferred.
Let $\mathcal{SP}\subseteq\mathcal R$ denote the set of all single-peaked preferences with respect to $<$.

The following characterization is the anonymous version of the classical result of \citet{Moulin1980}.
\begin{proposition}[\citet{Moulin1980,Weymark2011}]
A social choice function $F:\mathcal{SP}^n\to X$ is  anonymous, unanimous and strategy-proof if and only if there exists $\alpha = (\alpha_1, \ldots, \alpha_{n-1}) \in X^{n-1} \ (\alpha_1 \le \alpha_2 \le \ldots\le \alpha_{n-1})$ such that 
\[
F(R)=f^\alpha(\tau(R)) \ \text{ for every $R \in \mathcal{SP}^n$ }
\].
\end{proposition}
Anonymous and unanimous generalized median voter schemes therefore provide a natural benchmark for our analysis.
On the single-peaked domain, truthful reporting is a dominant strategy, so safe manipulation is impossible even under complete information.
In the analysis below, however, voters' preferences may be arbitrary elements of $\mathcal R$.
We investigate how much information about the other voters is required to manipulate these schemes safely once the single-peaked restriction is relaxed.

\subsubsection{RAT-degree}

We now characterize the RAT-degree of anonymous and unanimous generalized median voter schemes.
Although these schemes are strategy-proof on the single-peaked domain, they admit safe manipulation when preferences are unrestricted.
Rules in this class satisfy winner monotonicity.
Therefore, Theorem~\ref{thm:rat-upper-bound} implies that their
RAT-degree is at most $\lceil n/2\rceil$.

The following result shows that the minimum amount of information needed for such manipulation is determined by the numbers of fixed ballots at the two endpoints.

\begin{theorem}
\label{thm:rat-generalized-median}
Let $\alpha=(\alpha_1,\ldots,\alpha_{n-1})\in X^{n-1}$ satisfy $\alpha_1\le\cdots\le\alpha_{n-1}$.
On the unrestricted preference domain $\mathcal R$, the anonymous generalized median voter scheme $f^\alpha$ satisfies
\[
d_{\mathrm{RAT}}(f^\alpha)
=
1+\min\{\ell(\alpha),r(\alpha)\}.
\]
\end{theorem}

Theorem~\ref{thm:rat-generalized-median} provides an exact information threshold for safe manipulation.
If a voter knows the reports of at most $\min\{\ell(\alpha),r(\alpha)\}$ other voters, no safe manipulation is possible.
With one additional known report, there exist a preference relation and observed reports under which a safe manipulation becomes possible.

Only the smaller of the two endpoint counts matters for this threshold.
In particular, changing the locations of interior fixed ballots while preserving the endpoint counts leaves the RAT-degree unchanged.
If either endpoint has no fixed ballot, then $d_{\mathrm{RAT}}(f^\alpha)=1$: observing a single other voter's report can suffice for safe manipulation.

When $n$ is odd, the \emph{median voter rule} selects the median of the reported top alternatives:
\[
f^{\operatorname{med}}(t)= med(t_1,t_2,\ldots,t_n).
\]
This rule is the anonymous generalized median voter scheme obtained by placing $(n-1)/2$ fixed ballots at each endpoint:
\[
\alpha=
\bigl(
\underbrace{x_1,\ldots,x_1}_{(n-1)/2},
\underbrace{x_m,\ldots,x_m}_{(n-1)/2}
\bigr).
\]

For odd $n$, our characterization identifies the ordinal median rule as the most resistant to safe manipulation within the class of anonymous generalized median voter schemes, attaining the upper bound on the RAT-degree established in Theorem \ref{thm:rat-upper-bound}.
\begin{corollary}
\label{cor:rat-ordinal-median}
Suppose that $n$ is odd.
Then,
$d_{\mathrm{RAT}}(f^{\alpha})
=
\dfrac{n+1}{2}
=
\left\lceil\dfrac n2\right\rceil
$ if and only if $f^\alpha = f^\mathrm{med}.$
\end{corollary}

\subsection{Voting by quotas}

\subsubsection{Voting by quotas}

We now consider voting by quotas, the anonymous subclass of voting by committees introduced by \citet{Barbera1991}.
Let $O$ be a finite set of objects with $|O|\ge 2$, and let $X=2^O$.
Thus, each alternative is a subset of objects.
Let $\mathcal R_O$ denote the set of complete and transitive preference relations over $2^O$ with a unique top.
For each $R_i\in\mathcal R_O$, the top alternative $\tau(R_i)$ is voter $i$'s most preferred subset.

For a profile of reported top subsets $t=(t_i)_{i\in N}\in(2^O)^n$ and an object $a\in O$, let
\[
S_a(t)=\{i\in N:a\in t_i\}.
\]
Thus, $S_a(t)$ is the set of voters whose reported top subsets contain object $a$.
Let $q=(q_a)_{a\in O}\in\{1,\ldots,n\}^{O}$ be a vector of \emph{quotas}.
The \emph{voting by quotas} rule associated with $q$ is the tops-only voting rule $f^q:(2^O)^n\to 2^O$ defined by
\[
f^q(t)=\{a\in O:|S_a(t)|\ge q_a\}.
\]
Each object is selected if and only if the number of voters reporting it in their top subsets meets its quota.
Since selection depends only on these numbers, $f^q$ is anonymous.
Moreover, $q_a\in\{1,\ldots,n\}$ for every object ensures that $f^q$ is unanimous.

We next recall the classical connection between voting by quotas and strategy-proofness on the separable domain.
A preference relation $R_i\in\mathcal R_O$ is \emph{separable} if, for every $a\in O$ and every $B\subseteq O\setminus\{a\}$,
\[
\begin{aligned}
a\in\tau(R_i)
&\ \Longrightarrow\ (B\cup\{a\})P_i B,\\
a\notin\tau(R_i)
&\ \Longrightarrow\ B P_i(B\cup\{a\}).
\end{aligned}
\]
Thus, whether adding an object is desirable does not depend on which other objects are present.
Let $\mathcal S_O\subseteq\mathcal R_O$ denote the set of all separable preferences.

The following anonymous specialization follows from the characterization of voting by committees in \citet[Theorem~1]{Barbera1991}.

\begin{proposition}[\citet{Barbera1991}]
A social choice function $F:\mathcal S_O^n\to 2^O$ is anonymous, unanimous, and strategy-proof if and only if there exists $q\in\{1,\ldots,n\}^{O}$ such that
\[
F(R)=f^q(\tau(R))
\qquad
\text{for every }R\in\mathcal S_O^n.
\]
\end{proposition}

Voting by quotas therefore provides another natural benchmark for our analysis.
On the separable domain, truthful reporting is a dominant strategy, so safe manipulation is impossible even under complete information.
In the analysis below, however, voters' preferences may be arbitrary elements of $\mathcal R_O$.
We investigate how much information about the other voters is required to manipulate these rules safely once the separability restriction is relaxed.

\subsubsection{RAT-degree}

We now characterize the RAT-degree of voting by quotas.
The information threshold for safe manipulation is determined by the quotas for including objects and the corresponding thresholds for excluding them.

\begin{theorem}
\label{thm:rat-voting-by-quotas}
Let $q=(q_a)_{a\in O}\in\{1,\ldots,n\}^{O}$.
On the unrestricted preference domain $\mathcal R_O$, the voting by quotas rule $f^q$ satisfies
\[
d_{\mathrm{RAT}}(f^q)
=
\min_{a\in O}\min\{q_a,n-q_a+1\}.
\]
\end{theorem}

For each object $a$, inclusion is guaranteed by $q_a$ reports containing it, while exclusion is guaranteed by $n-q_a+1$ reports omitting it.
Theorem~\ref{thm:rat-voting-by-quotas} identifies the smallest of these thresholds across all objects as the RAT-degree.

When $n$ is odd, the \emph{objectwise majority rule} selects each object if and only if a majority of voters include it in their reported top subsets:
\[
f^{\mathrm{maj}}(t)
=
\left\{
a\in O:
|S_a(t)|\ge\frac{n+1}{2}
\right\}.
\]
This is the voting by quotas rule with $q_a=(n+1)/2$ for every $a\in O$.

For odd $n$, our characterization identifies objectwise majority as the unique voting by quotas rule most resistant to safe manipulation, attaining the upper bound on the RAT-degree established in Theorem~\ref{thm:rat-upper-bound}.

\begin{corollary}
\label{cor:rat-objectwise-majority}
Suppose that $n$ is odd.
Then,  $d_{\mathrm{RAT}}(f^q) = \dfrac{n+1}{2}= \left\lceil \dfrac {n}{2}\right\rceil$ if and only if $f^q = f^\mathrm{maj}$.

\end{corollary}

\section{Discussion}
\subsection{Majority criterion and optimality}
Theorem \ref{thm:rat-upper-bound} establishes an upper bound on the RAT-degree of anonymous, unanimous tops-only voting rules satisfying winner monotonicity.
We now identify a sufficient condition for attaining this bound.

The following axiom requires the selection of an alternative whenever it is the top choice of a strict majority of voters.

\paragraph{Majority criterion:}
A tops-only voting rule $f: X^n\to X$ satisfies the \textit{majority criterion} (MC) if, for every $t\in X^n$ and every $x\in X$, $|\{i\in N: t_i=x\}|>n/2$ implies $f(t)=x$.

Since the majority criterion implies unanimity, unanimity need not be imposed in the following theorem.

\begin{theorem}
\label{thm:optimal}
If a tops-only voting rule $f:X^n \to X$ satisfies anonymity, winner monotonicity, and the majority criterion, then
\[
d_{\mathrm{RAT}}(f) = \left\lceil \dfrac{n}{2}\right\rceil.
\]
\end{theorem}

This result provides a strategic rationale for the majority criterion: within this class, respecting strict majority support guarantees the largest possible information threshold for safe manipulation.

The majority criterion is sufficient but not necessary for attaining the upper bound, as shown by Example~\ref{counter_ex_thm2}.

\subsection{Comparison with existing measures of manipulability}
\label{subsec:comparison-manipulability}

\citet{ArribillagaMasso2016,ArribillagaMasso2017} compare generalized median voter schemes and voting by committees, respectively, through the sets of individual preferences that permit a profitable manipulation.
For a tops-only rule $f$, their criterion can be expressed in our notation using
\[
\mathcal M_i(f)
=
\left\{
R_i\in\mathcal R
\,\middle|\,
\begin{array}{l}
f(t_i',t_{-i})P_i f(\tau(R_i),t_{-i})\\
\text{for some }t_i'\in X
\text{ and }t_{-i}\in X^{n-1}
\end{array}
\right\}.
\]
A rule $f$ is at least as manipulable as another rule $g$ if $\mathcal M_i(g)\subseteq\mathcal M_i(f)$ for every voter $i$.
The two rules are equally manipulable if $\mathcal M_i(f)=\mathcal M_i(g)$ for every voter.
This criterion identifies the individual preferences under which a profitable deviation exists at some profile of the other voters' reports.
We now compare our results with \citet{ArribillagaMasso2016, ArribillagaMasso2017}.

\subsubsection{Generalized median voter schemes}
For odd $n\geq3$, the median voter rule illustrates the contrast between the two criteria.
Under the criterion of \citet[Corollary~5]{ArribillagaMasso2016}, it belongs to the most manipulable equivalence class of unanimous anonymous generalized median voter schemes.
Our characterization identifies the same rule as the unique maximizer of the RAT-degree within this class.

To examine this contrast, write $t_{(1)}\leq\cdots\leq t_{(n)}$ for the ordered reports and define $f^{(k)}(t)=t_{(k)}$ for each $k\in\{1,\ldots,n\}$.
All interior order-statistic rules belong to the most manipulable equivalence class under their criterion.
In particular, $\mathcal M_i(f^{(k)})=\mathcal M_i(f^{(k')})$ for every $i\in N$ and every $k,k'$ satisfying $1<k,k'<n$.
The same conclusion holds in our finite setting, including when preferences allow indifference below the unique top.\footnote{Fixing the other reports in any anonymous generalized median voter scheme gives a response to voter $i$'s report of the form $x\mapsto\operatorname{med}(x,u,v)$ for some $u\leq v$ in $X$.
Every such pair can be realized under any interior order-statistic rule $f^{(k)}$ by fixing $k-1$ other reports at $u$ and $n-k$ at $v$.
Thus, every profitable deviation under an anonymous generalized median voter scheme can be replicated under any interior order-statistic rule.}

The median voter rule leaves the entire set $X$ attainable conditional on any voter's report.
By comparison, narrowing the interval between the smallest and largest fixed ballots can restrict the set of attainable outcomes.
These broader options explain the median rule's position among the most manipulable rules under the criterion of \citet{ArribillagaMasso2016}.
The availability of profitable manipulation opportunities does not, however, imply that they can be exploited safely with little information.
Indeed, for odd $n$, safe manipulation of the median requires knowledge of at least $(n+1)/2$ other reports, attaining the maximal RAT-degree among unanimous anonymous generalized median voter schemes.

\subsubsection{Voting by quotas}
A parallel contrast arises for voting by quotas.
By \citet[Corollary~1]{ArribillagaMasso2017}, all quota rules satisfying $1<q_a<n$ for every $a\in O$ belong to the most manipulable equivalence class.
In particular, $\mathcal M_i(f^q)=\mathcal M_i(f^{\widehat q})$ for every $i\in N$ whenever both quota vectors satisfy this condition.
The same equivalence holds on our domain allowing indifference below the unique top.\footnote{For each object, an interior quota allows the other voters' reports to make the object always selected, always rejected, or selected precisely when voter $i$ reports it.
These possibilities are identical across interior quotas and can be arranged independently across objects.
Hence, the possible responses to voter $i$'s report are identical across all such quota rules.}
For odd $n\geq3$, this equivalence class contains the objectwise majority rule, defined by $q_a=(n+1)/2$ for every $a\in O$.
Our characterization identifies this rule as the unique maximizer of the RAT-degree among quota rules, attaining $(n+1)/2$.

As with generalized median voter schemes, the comparison can be interpreted through the alternatives attainable conditional on a voter's report.
When all quotas are interior, every subset of objects remains attainable under every report.
A quota of one allows a voter to guarantee an object's inclusion by reporting it, while a quota of $n$ allows her to guarantee its exclusion by omitting it.
These restrictions on attainable outcomes underlie the smaller sets of manipulable individual preferences associated with extreme quotas.
Indeed, \citet[Corollary~2]{ArribillagaMasso2017} show that rules satisfying $q_a\in\{1,n\}$ for every object admit no strictly less manipulable quota rule.

Such restrictions do not, however, guarantee resistance to safe manipulation.
By Theorem~\ref{thm:rat-voting-by-quotas}, even a single object with quota one or $n$ implies $d_{\mathrm{RAT}}(f^q)=1$.
Knowing one other report can fix that object's inclusion or exclusion, making the voter's top subset unattainable while leaving scope for a safe compromise involving another object.
Objectwise majority voting requires more information before such a guarantee can be established.
Thus, a quota rule can belong to the most manipulable equivalence class under their criterion and nevertheless require the greatest amount of information to manipulate safely.

\subsection{Relation to non-obvious manipulation}
\label{subsec:nom}
\citet{TroyanMorrill2020} introduce non-obvious manipulability (NOM), which rules out deviations that improve either the best possible outcome or the worst possible outcome relative to truthful reporting.
Both comparisons are evaluated according to the voter's true preference.
\citet{ArribillagaBonifacio2024} characterize NOM for tops-only voting rules and apply their characterization to generalized median voter schemes and voting by committees.
Their results provide a connection between NOM and our information thresholds for safe manipulation.

For each voter $i\in N$ and report $s_i\in X$, define the option set $\mathcal O_i(s_i)=\{f(s_i,t_{-i}):t_{-i}\in X^{n-1}\}$.
If $\mathcal O_i(s_i)=X$ for every voter and every report, then all reports leave the same best and worst outcomes attainable, so the rule is NOM.
This sufficient condition applies on our domain allowing indifference below the unique top.

For anonymous generalized median voter schemes, Theorem~\ref{thm:rat-generalized-median} implies that $d_{\mathrm{RAT}}(f^\alpha)\geq2$ if and only if $\ell(\alpha)\geq1$ and $r(\alpha)\geq1$.
Under this condition, every report leaves the entire set $X$ attainable.
Hence, $d_{\mathrm{RAT}}(f^\alpha)\geq2$ implies that $f^\alpha$ is NOM, consistently with the characterization of \citet{ArribillagaBonifacio2024}.
The converse need not hold.
For example, when $X=\{x_1,x_2,x_3\}$ and $\alpha=(x_2,\ldots,x_2)$, the scheme coincides with the unanimity rule with status quo $x_2$.
As discussed above, this rule is NOM but has RAT-degree one.

For voting by quotas, the connection is exact.
\citet{ArribillagaBonifacio2024} show that $f^q$ is NOM if and only if $2\leq q_a\leq n-1$ for every $a\in O$.
Together with Theorem~\ref{thm:rat-voting-by-quotas}, this yields that $f^q$ is NOM if and only if $d_{\mathrm{RAT}}(f^q)\geq2$.
Sufficiency continues to hold on our preference domain because interior quotas leave every subset of objects attainable under every report.
Necessity also continues to hold because their obvious manipulations at extreme quotas already exist under strict preferences.
The RAT-degree therefore further distinguishes quota rules that all satisfy NOM according to the amount of information required to manipulate them safely.

These implications do not extend to the entire class considered in Theorem~\ref{thm:rat-upper-bound}.
The voting rule introduced in Example \ref{counter_ex_thm2} attains the maximal RAT-degree of two but is not NOM.
For a voter with preference $aP_i bP_i c$, truthful reporting has worst possible outcome $c$, while reporting $b$ guarantees $b$.
The deviation therefore improves the worst possible outcome.
It remains risky, however, because when both other voters report $a$, it replaces the voter's top $a$ with $b$.
Thus, even an optimal RAT-degree need not imply NOM.

\begin{center}
\Large{{\bf Appendix}}
\end{center}

\numberwithin{definition}{section}
\numberwithin{theorem}{section}
\numberwithin{lemma}{section}
\numberwithin{proposition}{section}
\numberwithin{corollary}{section}
\numberwithin{example}{section}
\renewcommand{\theequation}{\thesection.\arabic{equation}}
\appendix

\section{Omitted Proofs}

\subsection{Proof of Theorem \ref{thm:rat-upper-bound}}

\begin{proof}
Let $h=\lceil n/2\rceil$ and $q=n-h=\lfloor n/2\rfloor$, and suppose that $d_{\mathrm{RAT}}(f)>h$.
Write $x^r$ for a block of $r$ reports equal to $x$; by anonymity, the order of the blocks is immaterial.
Fix $K\subseteq N$ with $|K|=h$ and $t_K\in X^K$, and let $H=N\setminus K$.
Define $\mathcal O(t_K)=\{f(t_K,s_H):s_H\in X^H\}$.

Suppose that $|\mathcal O(t_K)|\ge 2$ and $\mathcal{O}(t_K) \neq X.$ 
Choose $p\in X\setminus\mathcal O(t_K)$.
There exist $i\in H$, $s_i\in X$, and $\bar t_U\in X^U$, where $U=H\setminus\{i\}$, such that $f(p,t_K,\bar t_U)\neq f(s_i,t_K,\bar t_U)$.
Let $y=f(s_i,t_K,\bar t_U)$.
Choose $R_i\in\mathcal R$ such that $pP_i yP_i x$ and $xI_i z$ for all $x,z\in X\setminus\{p,y\}$.
Since $p\notin\mathcal O(t_K)$, we have $f(p,t_K,t_U)\neq p$ for every $t_U\in X^U$.
Moreover, winner monotonicity implies that $f(p,t_K,t_U)=y$ entails $f(y,t_K,t_U)=y$.
By the choice of $R_i$, it follows that $f(y,t_K,t_U)\,R_i\,f(p,t_K,t_U)$ for every $t_U\in X^U$.
At the specified completion, winner monotonicity gives $f(y,t_K,\bar t_U)=y$, and hence $f(y,t_K,\bar t_U)\,P_i\,f(p,t_K,\bar t_U)$.
Thus, the report $y$ is a safe manipulation given $(R_i,t_K)$.
Since $|K|=h$, this contradicts $d_{\mathrm{RAT}}(f)>h$.
Therefore, every restriction obtained by fixing $h$ reports has either a singleton range or range $X$.

Furthermore, if $\mathcal O(t_K)=X$, then $f(t_K,x_H)=x$ for every $x\in X$, where $x_H=(x)_{j\in H}$ by winner monotonicity.
By unanimity, the range obtained by fixing $h$ reports at $a$ contains $a$.
Thus, by anonymity and the preceding argument, either $h$ reports force $a$ or $f(a^h,x^q)=x$ for every $x\in X$.

Suppose that distinct alternatives $a,b\in X$ are both not forced by $h$ reports.
Then $f(a^h,b^q)=b$ and $f(a^q,b^h)=a$.
By winner monotonicity, $f(a^h,b^q)=b$ implies $f(a^{h-r},b^{q+r})=b$ for every $r\in\{0,\ldots,h-q\}$.
Taking $r=h-q$ gives $f(a^q,b^h)=b$.
This contradicts the second equality.
Hence, at most one alternative is not forced by $h$ reports.
Since $|X|\geq3$, there exist distinct $a,b\in X$ such that $h$ reports force $a$ and $h$ reports force $b$.
If $n=2q$, then $h=q$, so the profile $(a^q,b^q)$ must select both $a$ and $b$, a contradiction.
Suppose instead that $n=2q+1$, so that $h=q+1$.
Fix $q$ reports at $a$ and one report at $b$.
Since $2q\geq h$, completing the remaining reports with $a^q$ yields $a$, while completing them with $b^q$ yields $b$.
The range of this restriction therefore contains two distinct alternatives and must equal $X$.
Choose $c\in X\setminus\{a,b\}$.
The preceding observation then gives $f(a^q,b,c^q)=c$.
Let $\mathcal O(c^q,b)=\{f(c^q,b,s):s\in X^q\}$.
Since $q+1=h$ and $h$ reports force $b$, we have $f(c^q,b,b^q)=b$.
By anonymity, $f(c^q,b,a^q)=f(a^q,b,c^q)=c$.
Hence $\{b,c\}\subseteq\mathcal O(c^q,b)$, and since $b\neq c$, the preceding argument implies $\mathcal O(c^q,b)=X$.
Consequently, $f(c^q,b,a^q)=a$.
This contradicts anonymity and $f(a^q,b,c^q)=c$.
Therefore, $d_{\mathrm{RAT}}(f)\leq h=\lceil n/2\rceil$. 
\end{proof}

\subsection{Proof of Theorem \ref{thm:rat-generalized-median}}
\begin{proof}
Write $\ell=\ell(\alpha)$ and $r=r(\alpha)$.
We first show that $d_{\mathrm{RAT}}(f^\alpha)\geq 1+\min\{\ell,r\}$.
Fix any voter $i$, any set $K\subseteq N\setminus\{i\}$ with $k=|K|\leq\min\{\ell,r\}$, any $t_K\in X^K$, and any $R_i\in\mathcal R$.
Let $a=\tau(R_i)$, let $U=N\setminus(K\cup\{i\})$, and consider any misreport $b\neq a \ (b \in X)$.
By symmetry, it suffices to consider the case $b>a$.
Let $C:=|\{j\in\{1,\ldots,n-1\}:\alpha_j\leq a\}|+|\{j\in K:t_j\leq a\}|$ denote the number of ballots weakly below $a$ among the fixed ballots and the reports $t_K$.
Since $a<x_m$, we have $k\leq\ell\leq C\leq n-1-r+k\leq n-1$.
Hence, $0\leq n-1-C\leq n-1-k=|U|$, so we can choose $t_U$ such that exactly $n-1-C$ unknown voters report $a$ and all remaining unknown voters report $x_m$.
By construction, $|\{j\in\{1,\ldots,n-1\}:\alpha_j\leq a\}|+|\{j\in N\setminus\{i\}:t_j\leq a\}|=C+(n-1-C)=n-1$.
Truthful reporting therefore yields $f^\alpha(a,t_K,t_U)=a$, whereas reporting $b>a$ yields $f^\alpha(b,t_K,t_U)>a$.
Since $a$ is the unique top of $R_i$, we have $f^\alpha(a,t_K,t_U) P_i f^\alpha(b,t_K,t_U).$
Thus, no safe manipulation is possible when at most $\min\{\ell,r\}$ other voters’ reports are known, establishing the lower bound.

We next show that $d_{\mathrm{RAT}}(f^\alpha)\leq1+\min\{\ell,r\}$ by constructing a safe manipulation given the reports of $1+\min\{\ell,r\}$ other voters.
By symmetry, assume that $\ell\leq r$.
Let $k=\ell+1\leq n-1$ and fix a voter $i$.
Choose $R_i\in\mathcal R$ such that $x_1P_i x_mP_i x_{m-1}P_i\cdots P_i x_2$.
Choose a subset $S\subseteq N\setminus\{i\}$ with $|S|=r$, and define $\bar t_{-i}$ by setting $\bar t_j=x_2$ for every $j\in S$ and $\bar t_j=x_m$ for every $j\in N\setminus(S\cup\{i\})$.
Fix any $K\subseteq N\setminus\{i\}$ with $|K|=k$, and let $U=N\setminus(K\cup\{i\})$.
Define the observed report profile $t_K:=(\bar t_j)_{j\in K}\in X^K$.
We show that reporting $x_m$ is a safe manipulation given $(R_i,t_K)$.
Since $t_j>x_1$ for every $j\in K$, we have $|\{j\in K:t_j=x_1\}|=0$.
Hence, for every $t_U=(t_j)_{j\in U}\in X^U$, we have $|\{h\in\{1,\ldots,n-1\}:\alpha_h=x_1\}|+1+|\{j\in K:t_j=x_1\}|+|\{j\in U:t_j=x_1\}|=\ell+1+|\{j\in U:t_j=x_1\}|\leq\ell+1+|U|=n-1$, where $|U|=n-1-k=n-\ell-2$.
Hence, $f^\alpha(x_1,t_K,t_U)>x_1$ for every $t_U\in X^U$.
Since the median is weakly increasing in each report, we also have $f^\alpha(x_m,t_K,t_U)\geq f^\alpha(x_1,t_K,t_U)$ for every $t_U\in X^U$.
By the choice of $R_i$, these inequalities imply $f^\alpha(x_m,t_K,t_U)R_i f^\alpha(x_1,t_K,t_U)$ for every $t_U\in X^U$, establishing safety.
To establish profitability, consider the completion $t_U=\bar t_U$.
By construction, $|\{j\in\{1,\ldots,n-1\}:\alpha_j<x_m\}|+|\{j\in N\setminus\{i\}:\bar t_j<x_m\}|=(n-1-r)+r=n-1$.
Thus, truthful reporting yields $f^\alpha(x_1,t_K,\bar t_U)<x_m$, whereas reporting $x_m$ yields $f^\alpha(x_m,t_K,\bar t_U)=x_m$.
Since $f^\alpha(x_1,t_K,\bar t_U)>x_1$ and $x_m$ is ranked second under $R_i$, we obtain $f^\alpha(x_m,t_K,\bar t_U)P_i f^\alpha(x_1,t_K,\bar t_U)$.
Therefore, reporting $x_m$ is a safe manipulation given $(R_i,t_K)$, and $d_{\mathrm{RAT}}(f^\alpha)\leq\ell+1=1+\min\{\ell,r\}$.
Together with the lower bound established above, this completes the proof.
\end{proof}

\subsection{Proof of Theorem \ref{thm:rat-voting-by-quotas}}
\begin{proof}
Let $d=\min_{a\in O}\min\{q_a,n-q_a+1\}$.
Since $n\geq2$, we have $1\leq d\leq n-1$.
We first show that $d_{\mathrm{RAT}}(f^q)\geq d$.
Fix any voter $i$, any set $K\subseteq N\setminus\{i\}$ with $k=|K|<d$, any $t_K\in X^K$, and any $R_i\in\mathcal R_O$.
Let $U=N\setminus(K\cup\{i\})$ and $p=\tau(R_i)$.
For each $a\in O$, let $c_a=|\{j\in K:a\in t_j\}|$.
By the definition of $d$, we have $c_a\leq k<q_a$ and $q_a\leq n-k$.
Hence $0\leq q_a-1-c_a\leq n-1-k=|U|$.
For each $a\in O$, choose a set $U_a\subseteq U$ with $|U_a|=q_a-1-c_a$, and define $t_U=(t_j)_{j\in U}$ by $t_j=\{a\in O:j\in U_a\}$.
Under this completion, exactly $q_a-1$ voters other than $i$ report a subset containing $a$, for every $a\in O$.
Consequently, $f^q(s_i,t_K,t_U)=s_i$ for every report $s_i\in X$.
Thus, $f^q(p,t_K,t_U)=pP_is_i=f^q(s_i,t_K,t_U)$ for every $s_i\in X\setminus\{p\}$.
Thus no misreport is safe given $(R_i,t_K)$, establishing the lower bound.

We next show that $d_{\mathrm{RAT}}(f^q)\leq d$.
Choose an object $a\in O$ such that $d=\min\{q_a,n-q_a+1\}$, and choose another object $b\in O\setminus\{a\}$.
Fix a voter $i$ and a set $K\subseteq N\setminus\{i\}$ with $|K|=d$, and let $U=N\setminus(K\cup\{i\})$.
Choose a set $H\subseteq N\setminus\{i\}$ with $|H|=q_b-1$.
Suppose first that $d=q_a$.
Define $t_K=(t_j)_{j\in K}$ by $t_j=\{a,b\}$ for $j\in K\cap H$ and $t_j=\{a\}$ for $j\in K\setminus H$.
Choose $R_i\in\mathcal R_O$ such that $\varnothing P_i\{a,b\}P_i x$ and $xI_i y$ for all $x,y\in X\setminus\{\varnothing,\{a,b\}\}$, where $I_i$ denotes indifference under $R_i$.
Consider the misreport $s_i=\{b\}$.
Fix any $t_U\in X^U$, and let $x=f^q(\varnothing,t_K,t_U)$ and $y=f^q(\{b\},t_K,t_U)$.
Since $|\{j\in K:a\in t_j\}|=|K|=q_a$, we have $a\in x\cap y$, and hence $x,y\neq\varnothing$.
By the definition of $f^q$, we also have $y\in\{x,x\cup\{b\}\}$.
If $x=\{a,b\}$, then $y=x$.
If $x\neq\{a,b\}$, then $x\in X\setminus\{\varnothing,\{a,b\}\}$, so the definition of $R_i$ implies $yR_i x$.
Consequently, $f^q(\{b\},t_K,t_U)R_if^q(\varnothing,t_K,t_U)$ for every $t_U\in X^U$.
To establish profitability, define $t_U=(t_j)_{j\in U}$ by $t_j=\{b\}$ for $j\in U\cap H$ and $t_j=\varnothing$ for $j\in U\setminus H$.
Under this completion, exactly $q_a$ other voters report $a$, exactly $q_b-1$ other voters report $b$, and no voter reports any other object.
Hence  $f^q(\{b\},t_K,t_U)=\{a,b\}P_i\{a\}=f^q(\varnothing,t_K,t_U).$
Thus, $s_i$ is a safe manipulation given $(R_i,t_K)$.
The case $d=n-q_a+1$ follows by a symmetric argument.
Combining the two bounds gives $d_{\mathrm{RAT}}(f^q)=d$, which completes the proof.
\end{proof}

\subsection{Proof of Theorem \ref{thm:optimal}}
\begin{proof}
Let $h=\lceil n/2\rceil$.
Since the majority criterion implies unanimity, Theorem \ref{thm:rat-upper-bound} gives $d_{\mathrm{RAT}}(f)\leq h$.
We establish the reverse inequality by showing that every misreport can be harmful whenever fewer than $h$ reports are known.
Fix any voter $i\in N$, any $K\subseteq N\setminus\{i\}$ with $k=|K|<h$, any $t_K\in X^K$, and any $R_i\in\mathcal R$.
Let $U=N\setminus(K\cup\{i\})$, let $p=\tau(R_i)$, and fix any misreport $s_i\neq p$.
Set $m=n-k=|U|+1$.
Write $x^r$ for a block of $r$ reports equal to $x$, with $x^0$ denoting an empty block; anonymity allows us to disregard the order of the reports.
Since $m>n/2$, the majority criterion gives $f(t_K,p^m)=p$ and $f(t_K,s_i^m)=s_i$.
Let $r$ be the smallest integer satisfying $1\leq r\leq m$ such that $f(t_K,p^{m-r},s_i^r)\neq p$.
Such an integer exists because $f(t_K,s_i^m)=s_i\neq p$, and its minimality implies $f(t_K,p^{m-r+1},s_i^{r-1})=p$.
Choose $B\subseteq U$ with $|B|=r-1$, which is possible because $0\leq r-1\leq m-1=|U|$.
Define $\bar t_U=(\bar t_j)_{j\in U}$ by $\bar t_j=s_i$ for $j\in B$ and $\bar t_j=p$ for $j\in U\setminus B$.
By anonymity, truthful reporting yields $f(p,t_K,\bar t_U)=f(t_K,p^{m-r+1},s_i^{r-1})=p$.
The misreport instead yields $f(s_i,t_K,\bar t_U)=f(t_K,p^{m-r},s_i^r)\neq p$.
Since $p$ is the unique top of $R_i$, we have $f(p,t_K,\bar t_U)\,P_i\,f(s_i,t_K,\bar t_U)$.
Thus, $s_i$ is not safe given $(R_i,t_K)$.
Since all choices were arbitrary, no safe manipulation is possible with fewer than $h$ known reports, so $d_{\mathrm{RAT}}(f)\geq h$.
Combining the two bounds gives $d_{\mathrm{RAT}}(f)=\lceil n/2\rceil$. 
\end{proof}

\bibliography{references_RAT}

@article{Moulin1980,
  author  = {Moulin, Herv{\'e}},
  title   = {On Strategy-Proofness and Single-Peakedness},
  journal = {Public Choice},
  year    = {1980},
  volume  = {35},
  number  = {4},
  pages   = {437--455},
  doi     = {10.1007/BF00128122}
}

@inproceedings{Hartman2025,
  author    = {Hartman, Eden and Segal-Halevi, Erel and Tao, Biaoshuai},
  title     = {It's Not All Black and White: Degree of Truthfulness for Risk-Avoiding Agents},
  booktitle = {Proceedings of the 26th ACM Conference on Economics and Computation},
  year      = {2025},
  pages     = {996--1016},
  publisher = {Association for Computing Machinery},
  doi       = {10.1145/3736252.3742658}
}

@article{Bu2023,
  author  = {Bu, Xiaolin and Song, Jiaxin and Tao, Biaoshuai},
  title   = {On Existence of Truthful Fair Cake Cutting Mechanisms},
  journal = {Artificial Intelligence},
  year    = {2023},
  volume  = {319},
  pages   = {103904},
  doi     = {10.1016/j.artint.2023.103904}
}

@article{Muller2024,
  author  = {M{\"u}ller, Michael},
  title   = {Belief-Independence and (Robust) Strategy-Proofness},
  journal = {Theory and Decision},
  year    = {2024},
  volume  = {96},
  number  = {3},
  pages   = {443--461},
  doi     = {10.1007/s11238-023-09955-7}
}

@article{Barbera1991,
  author  = {Barber{\`a}, Salvador and Sonnenschein, Hugo and Zhou, Lin},
  title   = {Voting by Committees},
  journal = {Econometrica},
  year    = {1991},
  volume  = {59},
  number  = {3},
  pages   = {595--609},
  doi     = {10.2307/2938220}
}

@article{ArribillagaMasso2016,
  author  = {Arribillaga, R. Pablo and Mass{\'o}, Jordi},
  title   = {Comparing Generalized Median Voter Schemes According to Their Manipulability},
  journal = {Theoretical Economics},
  year    = {2016},
  volume  = {11},
  number  = {2},
  pages   = {547--586},
  doi     = {10.3982/TE1910}
}

@article{ArribillagaMasso2017,
  author  = {Arribillaga, R. Pablo and Mass{\'o}, Jordi},
  title   = {Comparing Voting by Committees According to Their Manipulability},
  journal = {American Economic Journal: Microeconomics},
  year    = {2017},
  volume  = {9},
  number  = {4},
  pages   = {74--107},
  doi     = {10.1257/mic.20160107}
}

@article{MausPetersStorcken2007,
  author  = {Maus, Stefan and Peters, Hans and Storcken, Ton},
  title   = {Anonymous Voting and Minimal Manipulability},
  journal = {Journal of Economic Theory},
  year    = {2007},
  volume  = {135},
  number  = {1},
  pages   = {533--544},
  doi     = {10.1016/j.jet.2006.04.002}
}

@article{ArribillagaBonifacio2024,
  author  = {Arribillaga, R. Pablo and Bonifacio, Agust{\'i}n G.},
  title   = {Obvious Manipulations of Tops-Only Voting Rules},
  journal = {Games and Economic Behavior},
  year    = {2024},
  volume  = {143},
  pages   = {12--24},
  doi     = {10.1016/j.geb.2023.11.002}
}

@article{TroyanMorrill2020,
  author  = {Troyan, Peter and Morrill, Thayer},
  title   = {Obvious Manipulations},
  journal = {Journal of Economic Theory},
  year    = {2020},
  volume  = {185},
  pages   = {104970},
  doi     = {10.1016/j.jet.2019.104970}
}

@article{Gibbard1973,
  author  = {Gibbard, Allan},
  title   = {Manipulation of Voting Schemes: A General Result},
  journal = {Econometrica},
  year    = {1973},
  volume  = {41},
  number  = {4},
  pages   = {587--601},
  doi     = {10.2307/1914083}
}

@article{Satterthwaite1975,
  author  = {Satterthwaite, Mark Allen},
  title   = {Strategy-Proofness and {Arrow}'s Conditions: Existence and Correspondence Theorems for Voting Procedures and Social Welfare Functions},
  journal = {Journal of Economic Theory},
  year    = {1975},
  volume  = {10},
  number  = {2},
  pages   = {187--217},
  doi     = {10.1016/0022-0531(75)90050-2}
}

@inproceedings{ConitzerWalshXia2011,
  author    = {Conitzer, Vincent and Walsh, Toby and Xia, Lirong},
  title     = {Dominating Manipulations in Voting with Partial Information},
  booktitle = {Proceedings of the Twenty-Fifth AAAI Conference on Artificial Intelligence},
  year      = {2011},
  pages     = {638--643},
  publisher = {AAAI Press},
  doi       = {10.1609/aaai.v25i1.7879}
}

@inproceedings{ReijngoudEndriss2012,
  author    = {Reijngoud, Annemieke and Endriss, Ulle},
  title     = {Voter Response to Iterated Poll Information},
  booktitle = {Proceedings of the 11th International Conference on Autonomous Agents and Multiagent Systems},
  year      = {2012},
  volume    = {2},
  pages     = {635--644},
  publisher = {International Foundation for Autonomous Agents and Multiagent Systems},
  url       = {https://ifaamas.org/Proceedings/aamas2012/papers/2D_5.pdf}
}

@inproceedings{EndrissEtAl2016,
  author    = {Endriss, Ulle and Obraztsova, Svetlana and Polukarov, Maria and Rosenschein, Jeffrey S.},
  title     = {Strategic Voting with Incomplete Information},
  booktitle = {Proceedings of the Twenty-Fifth International Joint Conference on Artificial Intelligence},
  year      = {2016},
  pages     = {236--242},
  publisher = {AAAI Press},
  url       = {https://www.ijcai.org/Abstract/16/041}
}

@article{Klamler2014,
  author  = {Klamler, Christian},
  title   = {How Risky Is It to Manipulate a Scoring Rule under Incomplete Information?},
  journal = {Economics Bulletin},
  year    = {2014},
  volume  = {34},
  number  = {2},
  pages   = {1214--1221},
  url     = {https://accessecon.com/includes/CountdownloadPDF.aspx?PaperID=EB-14-00486}
}

@article{Weymark2011,
  author  = {Weymark, John A.},
  title   = {A Unified Approach to Strategy-Proofness for Single-Peaked Preferences},
  journal = {SERIEs},
  year    = {2011},
  volume  = {2},
  number  = {4},
  pages   = {529--550},
  doi     = {10.1007/s13209-011-0064-5}
}

@article{LeBretonSen1999,
  author  = {Le Breton, Michel and Sen, Arunava},
  title   = {Separable Preferences, Strategyproofness, and Decomposability},
  journal = {Econometrica},
  year    = {1999},
  volume  = {67},
  number  = {3},
  pages   = {605--628},
  doi     = {10.1111/1468-0262.00038}
}

@article{DryzekList2003,
  author  = {Dryzek, John S. and List, Christian},
  title   = {Social Choice Theory and Deliberative Democracy: A Reconciliation},
  journal = {British Journal of Political Science},
  year    = {2003},
  volume  = {33},
  number  = {1},
  pages   = {1--28},
  doi     = {10.1017/S0007123403000012}
}

@article{Nurmi2023,
  author  = {Nurmi, Hannu},
  title   = {Deliberative Democracy and Incompatibilities of Choice Norms},
  journal = {Behavioral Sciences},
  year    = {2023},
  volume  = {13},
  number  = {12},
  pages   = {985},
  doi     = {10.3390/bs13120985}
}

@unpublished{Fernandez2020,
  author = {Fernandez, Marcelo Ariel},
  title  = {Deferred Acceptance and Regret-Free Truth-Telling},
  year   = {2020},
  note   = {Working paper},
  url    = {https://jhir.library.jhu.edu/handle/1774.2/65832}
}

@article{ChenMoller2024,
  author  = {Chen, Yiqiu and M{\"o}ller, Markus},
  title   = {Regret-Free Truth-Telling in School Choice with Consent},
  journal = {Theoretical Economics},
  year    = {2024},
  volume  = {19},
  number  = {2},
  pages   = {635--666},
  doi     = {10.3982/TE4979}
}

@unpublished{ArribillagaBonifacioFernandez2022,
  author = {Arribillaga, R. Pablo and Bonifacio, Agust{\'i}n G. and Fernandez, Marcelo Ariel},
  title  = {Regret-Free Truth-Telling Voting Rules},
  year   = {2022},
  note   = {Working paper, arXiv:2208.13853},
  url    = {https://arxiv.org/abs/2208.13853}
}

\end{document}